\documentstyle[12pt,epsfig]{article}

\advance\textheight by \topskip
\begin{document}
\tolerance=100000 

\newcommand{\be}{\begin{equation}} 
\newcommand{\ee}{\end{equation}}
\newcommand{\br}{\begin{eqnarray}} 
\newcommand{\er}{\end{eqnarray}}
\newcommand{\ba}{\begin{array}} 
\newcommand{\ea}{\end{array}}
\newcommand{\bi}{\begin{itemize}} 
\newcommand{\ei}{\end{itemize}}
\newcommand{\bn}{\begin{enumerate}} 
\newcommand{\en}{\end{enumerate}}
\newcommand{\bc}{\begin{center}} 
\newcommand{\ec}{\end{center}}
\newcommand{\ul}{\underline} 
\newcommand{\ol}{\overline}
\def\epem{\ifmmode{e^+ e^-} \else{$e^+ e^-$} \fi}
\newcommand{\eeww}{$e^+e^-\rightarrow W^+ W^-$}
\newcommand{\qqQQ}{$q_1\bar q_2 Q_3\bar Q_4$}
\newcommand{\eeqqQQ}{$e^+e^-\rightarrow q_1\bar q_2 Q_3\bar Q_4$}
\newcommand{\eewwqqqq}{$e^+e^-\rightarrow W^+ W^-\ar q\bar q Q\bar Q$}
\newcommand{\eeqqgg}{$e^+e^-\rightarrow q\bar q gg$}
\newcommand{\eeqloop}{$e^+e^-\rightarrow q\bar q gg$ via quark loops}
\newcommand{\eeqqqq}{$e^+e^-\rightarrow q\bar q Q\bar Q$}
\newcommand{\eewwjjjj}{$e^+e^-\rightarrow W^+ W^-\rightarrow
  4~{\rm{jet}}$} 
\newcommand{\eeqqggjjjj}{$e^+e^-\rightarrow q\bar q
  gg\rightarrow 4~{\rm{jet}}$}
\newcommand{\eeqloopjjjj}{$e^+e^-\rightarrow q\bar q gg\rightarrow
  4~{\rm{jet}}$ via quark loops}
\newcommand{\eeqqqqjjjj}{$e^+e^-\rightarrow q\bar q Q\bar Q\rightarrow
  4~{\rm{jet}}$} 
\newcommand{\eejjjj}{$e^+e^-\rightarrow 4~{\rm{jet}}$} 
\newcommand{\jjjj}{$4~{\rm{jet}}$}
\newcommand{\qqbar}{$q\bar q$} 
\newcommand{\ww}{$W^+W^-$}
\newcommand{\ar}{\rightarrow} 
\newcommand{\sm}{${\cal {SM}}$}
\newcommand{\Dir}{\kern -6.4pt\Big{/}} 
\newcommand{\Dirin}{\kern
  -10.4pt\Big{/}\kern 4.4pt} 
\newcommand{\DDir}{\kern -7.6pt\Big{/}}
\newcommand{\DGir}{\kern -6.0pt\Big{/}} 
\newcommand{\wwqqqq}{$W^+W^-\ar q\bar q Q\bar Q$} 
\newcommand{\qqgg}{$q\bar q gg$}
\newcommand{\qloop}{$q\bar q gg$ via quark loops}
\newcommand{\qqqq}{$q\bar q Q\bar Q$} 
\newcommand{\ord}{{\cal O}}
\newcommand{\Ecm}{E_{\mathrm{cm}}}

\def\l{\left\langle}
\def\r{\right\rangle}
\def\aem{\alpha_{\rm em}}
\def\as{\alpha_{\rm s}}
\def\MW{M_{W^\pm}}
\def\MZ{M_{Z}}
\def\ycut{y_{\rm cut}}
\def\Ord{\lower .7ex\hbox{$\;\stackrel{\textstyle <}{\sim}\;$}}
\def\OOrd{\lower .7ex\hbox{$\;\stackrel{\textstyle >}{\sim}\;$}}
\def\pl #1 #2 #3 {{\it Phys.~Lett.} {\bf#1} (#2) #3}
\def\np #1 #2 #3 {{\it Nucl.~Phys.} {\bf#1} (#2) #3}
\def\jp #1 #2 #3 {{\it J.~Phys.} {\bf#1} (#2) #3}
\def\zp #1 #2 #3 {{\it Z.~Phys.} {\bf#1} (#2) #3}
\def\pr #1 #2 #3 {{\it Phys.~Rev.} {\bf#1} (#2) #3}
\def\prep #1 #2 #3 {{\it Phys.~Rep.} {\bf#1} (#2) #3}
\def\prl #1 #2 #3 {{\it Phys.~Rev.~Lett.} {\bf#1} (#2) #3}
\def\mpl #1 #2 #3 {{\it Mod.~Phys.~Lett.} {\bf#1} (#2) #3}
\def\rmp #1 #2 #3 {{\it Rev. Mod. Phys.} {\bf#1} (#2) #3}
\def\sjnp #1 #2 #3 {{\it Sov. J. Nucl. Phys.} {\bf#1} (#2) #3}
\def\cpc #1 #2 #3 {{\it Comp. Phys. Commun.} {\bf#1} (#2) #3}
\def\xx #1 #2 #3 {{\bf#1}, (#2) #3}
\def\preprint{{\it preprint}}

\begin{flushright}
{\large DESY-00-011}\\
{\large RAL-TR-1999-073}\\
{\rm Nov 1999\hspace*{.5 truecm}}\\
\end{flushright}

\vspace{0.1cm}

\begin{center}
  {\Large \bf The triple Higgs self-coupling at future $e^+e^-$
    colliders: a signal-to-background study for the standard
    model\footnote{Talk given at the Fourth Workshop of the {\it ``2nd
        Joint ECFA/DESY Study on Physics and Detectors for a Linear
        Electron-Positron
        Collider,''} Oxford, UK, 20-23 March 1999.}}\\[0.4cm]
  D.J. MILLER$^{1}$ and S.~MORETTI$^{2}$\\[0.4 cm]
  {\it 1) Deutches Elektronen-Synchrotron DESY, D-22603 Hamburg, Germany}\\
  {\it 2) Rutherford Appleton Laboratory, Chilton, Didcot, Oxon OX11 0QX, UK.}\\
\end{center}
\begin{abstract}
{\small
\noindent
The experimental reconstruction of the Higgs self-energy potential is
essential to a verification of the Higgs boson's r\^ole in spontaneous
electroweak symmetry breaking. The first step towards this goal, the
measurement of the triple Higgs self-coupling, can possibly be
accomplished at the next generation of linear colliders.  Here we
present a background study of the most promising channel, double
Higgs-strahlung off a $Z$ boson, $e^+e^-\to HHZ$, with the subsequent
decay $H\ar b\bar b$, and evaluate the feasibility of its measurement.}
\end{abstract}

\section{Introduction}
\label{sec_intro}

To verify whether or not the Higgs mechanism is responsible for
spontaneous electroweak symmetry breaking as expected by the standard
model {\sc (Sm)}, or indeed, the minimal supersymmetric standard model
{\sc (Mssm)}, one must perform an experimental reconstruction of the
Higgs self-energy potential.  This reconstruction requires the
measurement of the Higgs boson mass(es) and the triple and quartic
Higgs self-couplings.  The triple Higgs self-coupling may become
accessible to direct measurement at an electron-positron linear
collider {\sc (Lc)} with centre-of-mass energy at the TeV scale. In
this study we compare the signal of the most promising channel with
its dominant electroweak {\sc (Ew)} and {\sc Qcd} backgrounds to
assess the feasibility of its measurement.

We restrict ourselves to a discussion of the triple Higgs
self-coupling in the {\sc Sm}\footnote{Some of the trilinear couplings
  of the {\sc Mssm} may prove to be accessible even at the {\sc Lhc}
  via resonant decay of a heavy Higgs, eg. $H \to hh$, and are under
  investigation elsewhere \cite{MMM}.} and in particular double
Higgs-strahlung off $Z$ bosons, in the process $e^+e^-\to HHZ$.  We
adopt the $H \to b\bar b$ decay channel over the Higgs mass range
$M_H\Ord140$ GeV and assume very efficient tagging and high-purity
sampling of $b$ quarks. Then the backgrounds to the $\lambda_{HHH}$
measurement are primarily the `irreducible' {\sc Ew} and {\sc Qcd}
backgrounds $e^+e^- \to b \bar b b \bar b Z$.

\begin{figure}[ht]
\begin{center}
\psfig{file=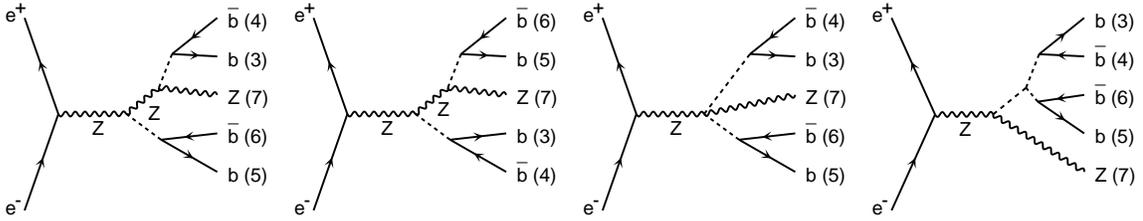,height=2.3cm,bbllx=110pt,bblly=665pt,
                                bburx=665pt,bbury=755pt}
\caption{Diagrams contributing at lowest-order to
  $e^+ e^- \to b_{(3)} \bar b_{(4)} b_{(5)} \bar b_{(6)}
  Z_{(7)}$ via purely {\sc Ew} interactions containing two Higgs
  bosons in the intermediate state. Note that only the far right
  diagram contains a dependence on $\lambda_{HHH}$.}
\end{center}
\label{fig_HH}
\end{figure}

The {\sc Ew} background is of $\ord(\alpha_{em}^5)$ away from
resonances, but can, in principle, be problematic due to the presence
of both $Z$ vectors and Higgs scalars yielding $b \bar b$ pairs.  In
contrast, the {\sc Qcd} background is of $\ord(\alpha_{em}^3
\alpha_s^2)$. Here, although there are no heavy objects decaying to $b
\bar b$ pairs, the production rate itself could give difficulties due
to the presence of the strong coupling.  In addition, the double
Higgs-strahlung process (see Fig.~\ref{fig_HH}) contains diagrams
proceeding via an $HHZ$ intermediate state but not dependent on
$\lambda_{HHH}$, as well as a diagram sensitive to the triple Higgs
self-coupling (the right-hand graph of Fig.~\ref{fig_HH}).

The plan of the paper is as follows. The next section details the
procedure adopted in computing the relevant scattering amplitudes.
Section~\ref{sec_results} displays our numerical results and contains
our discussion. Finally, in the last section, we summarize and
conclude.

\section{The calculation}
\label{sec_calculation}

The signal process is rather straightforward to calculate in the case
of on-shell $HHZ$ production (see Refs.~\cite{GlMa,PMZ} for analytic
expressions of the matrix elements ({\sc Me}s) and further discussions
of related signal processes).  The {\sc Ew} background is more
complex, deriving from many graphs with different resonant structures.
In order to perform an efficient integration we have grouped the
Feynman diagrams into different collections of diagrams with identical
(non-)resonant structure.  This categorization allows one to compute
each of the topologies separately, with the appropriate mapping of
variables, thus optimizing the accuracy of the numerical integration.
Furthermore, one is able to assess the relative weight of the various
subprocesses in the full scattering amplitude, giving added insight
into the fundamental dynamics.  In contrast the {\sc Qcd} background
contains fewer diagrams, with only five different (non-)resonant
topologies. This makes integration much simpler than in the {\sc Ew}
case and it can be carried out with percent accuracy directly over the
full {\sc Me} using standard multichannel Monte Carlo methods. For
further details and numerical inputs see Ref.~\cite{US}.

We assume that total and differential
rates are those at the partonic level, as we identify jets with the
partons from which they originate.  To resolve the final
state $b$ (anti)quarks as separate systems, we impose the following
acceptance cuts: $E(b)>10$ GeV on the energy of each $b$ (anti)quark
and $\cos(b,b)<0.95$ on the relative separation of all possible $2b$
combinations. We further assume that $b$ jets are distinguishable from
light-quark and gluon jets (by using, for example, $\mu$-vertex
tagging techniques). However, no efficiency to tag the four $b$ quarks
is included in our results.  The $Z$ boson is treated as
on-shell and no branching ratio is applied to quantify its
possible decays. In practice, in order to simplify the treatment of
the final state, one may assume that the $Z$ boson decays leptonically
(that is, $Z\to \ell^+\ell^-$, with $\ell=e,\mu,\tau$) or hadronically
into light quark jets (that is, $Z\to q\bar q$, with $q\ne b$).
Also, we have not included Initial State Radiation {\sc (Isr)}
\cite{ISR} in our calculations.  
 
\section{Results}
\label{sec_results}

\begin{figure}[ht]
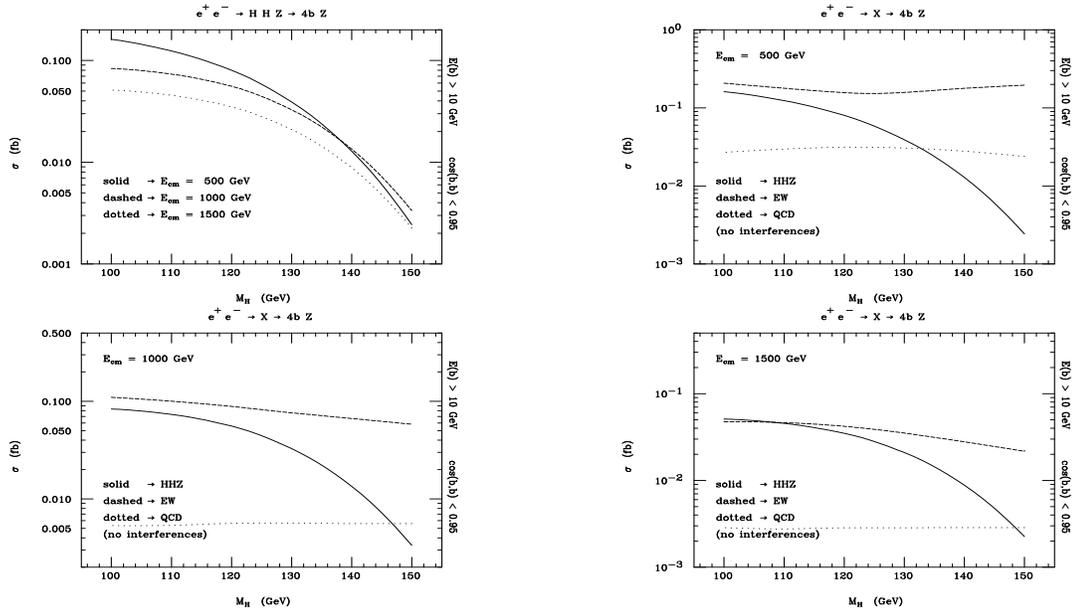

\begin{minipage}[b]{.495\linewidth}
\centering\psfig{file=signal.ps,angle=90,width=6cm,height=4cm} 
\end{minipage}\hfill\hfill
\begin{minipage}[b]{.495\linewidth}
\centering\psfig{file=background500.ps,angle=90,width=6cm,height=4cm}
\end{minipage}\hfill\hfill
\begin{minipage}[b]{.495\linewidth}
\centering\psfig{file=background1000.ps,angle=90,width=6cm,height=4cm}
\end{minipage}\hfill\hfill
\begin{minipage}[b]{.495\linewidth}
\centering\psfig{file=background1500.ps,angle=90,width=6cm,height=4cm}
\end{minipage}
\caption{Top-left: cross sections in femtobarns for the signal at three 
  different collider energies: 500, 1000 and 1500 GeV.
  Top-right(Bottom-left)[Bottom-right]: cross sections in femtobarns
  for the signal versus the {\sc Ew} and {\sc Qcd} backgrounds at
  500(1000)[1500] GeV.  Our acceptance cuts in energy and separation
  of the four $b$ quarks have been implemented.}
\label{fig_cross}
\end{figure}

The total signal cross section is plotted as a function of $M_H$ in
the top-left frame of Fig.~\ref{fig_cross}, for three centre-of-mass
{\sc (Cm)} energies.  Even at low Higgs masses where both the
production and decay rates are largest, the signal is rather small. In
fact, the signal is below $0.2$ femtobarns for all energies from 500
to 1500 GeV, although this can be doubled simply by polarizing the
incoming electron and positron beams.  Thus, as already recognized in
Ref.~\cite{PMZ}, where on-shell production studies of the signal were
performed, luminosities of the order of one inverse attobarn need to
be collected before statistically significant measurements of
$\lambda_{HHH}$ can be performed. 

The decrease of the signal with increasing Higgs mass is due to
suppression of the $H\to b\bar b$ decay channel. Above $M_H\approx140$
GeV it becomes overwhelmed by the opening of the off-shell $H\ar
W^{\pm*}W^{\mp}$ decay (see, for example, Fig.~1 of
Ref.~\cite{WJSZK}). In contrast, the production cross section for
$e^+e^-\to HHZ$ without specifying the subsequent decay is much less
sensitive to $M_H$ \cite{PMZ}.  In addition, because the signal is an
annihilation process proportional to $1/s$, a larger {\sc Cm} energy
($E_{\mathrm{cm}}$) tends to deplete the production rates, as long as
$E_{\mathrm{cm}}\gg 2M_H+M_Z$. When this is no longer true, e.g., at
500 GeV and $M_H\OOrd140$ GeV, phase space suppression can overturn
the $1/s$ propagator effects. This is evidenced by the crossing of the
curves for 500 and 1000 GeV in Fig.~\ref{fig_cross}.

\begin{figure}[ht]
\begin{center}
\begin{minipage}{7.5cm}
\psfig{figure=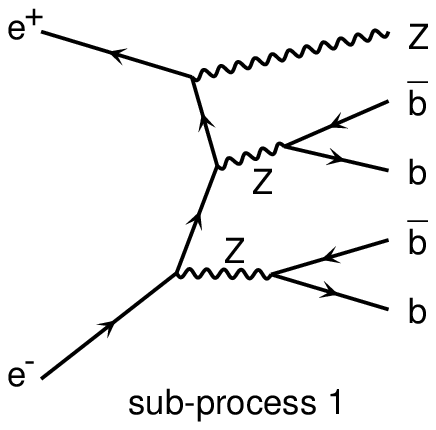,width=3.3cm,bbllx=85pt,bblly=483pt,
                                bburx=215pt,bbury=605pt}
\psfig{figure=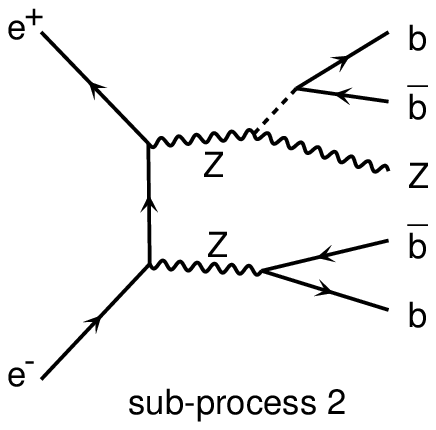,width=3.3cm,bbllx=101pt,bblly=33pt,
                                bburx=231pt,bbury=155pt} 
\psfig{figure=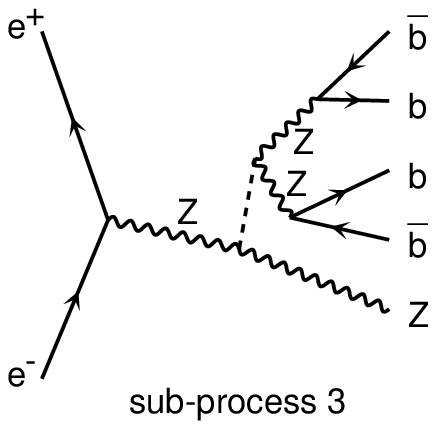,width=3.3cm,bbllx=385pt,bblly=333pt,
                                bburx=515pt,bbury=455pt}
\psfig{figure=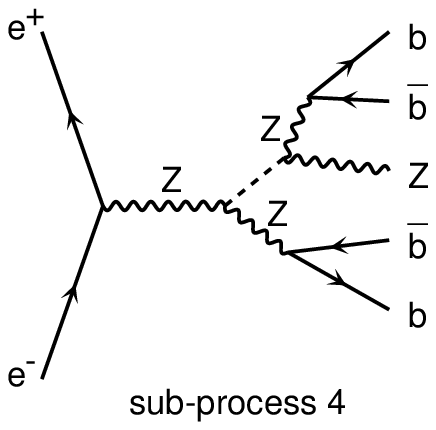,width=3.3cm,bbllx=372pt,bblly=633pt,
                                bburx=502pt,bbury=744pt}
\end{minipage}
\begin{minipage}{7.5cm}
\psfig{file=split_EW.ps,width=6cm,height=7cm,angle=90} 
\end{minipage}
\caption{On the left, generic Feynman diagrams representing 
  the four dominant {\sc Ew} topologies.  Note that only sub-process
  $1$ does not contain a Higgs boson in the intermediate state. The
  different contributions of these sub-processes to the total {\sc Ew}
  background are displayed in the plot to the right, as a function of
  the Higgs boson mass.}
\label{fig_H}
\end{center}
\end{figure}

Fig.~\ref{fig_cross} also shows the background processes plotted with
respect to $M_H$ for the three {\sc Cm} energies. As anticipated, the
{\sc Ew} background is problematic due to its resonance structures,
whereas the {\sc Qcd} background presents no such difficulty and does
not significantly obscure the signal, despite the strong coupling
constant.  Our categorization of the {\sc Ew} background into
different resonant topologies now facilitates a closer examination.
In particular we observe that only four sub-processes dominate the
background.  Generic Feynman diagrams of these sub-processes can be
seen in Fig.~\ref{fig_H}, together with their contribution to the
total {\sc Ew} background rate, as a function of the Higgs boson mass.
All other {\sc Ew} sub-processes are much smaller, rarely exceeding
$10^{-3}$ femtobarns, and have little effect on the kinematics or
magnitude of the background.

The {\sc Qcd} background is dominated by $e^+e^-\ar ZZ$ production
with one of the two $Z$ bosons decaying hadronically into four $b$
jets.  Also significant is single Higgs-strahlung production (off a
$Z$) with the Higgs scalar subsequently decaying into $b\bar bb\bar b$
via an off-shell gluon. The contributions of the other diagrams, which
do not resonate, are typically one order of magnitude smaller than the
$ZZ$ and $ZH$ mediated graphs, with the interferences smaller still
(and generally negative).

\begin{figure}[ht]
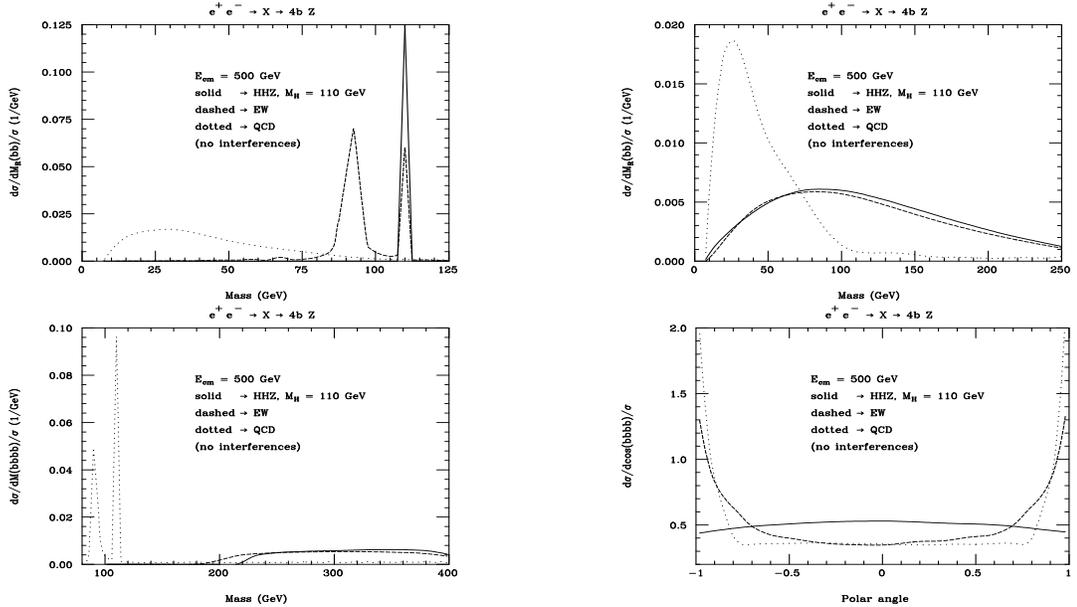

\begin{minipage}[b]{.495\linewidth}
\centering\epsfig{file=mbbR_procs.ps,width=4cm,height=6cm,angle=90}
\end{minipage}\hfill\hfill
\begin{minipage}[b]{.495\linewidth}
\centering\epsfig{file=mbbW_procs.ps,width=4cm,height=6cm,angle=90}
\end{minipage}\hfill\hfill
\begin{minipage}[b]{.495\linewidth}
\centering\epsfig{file=mbbbb_procs.ps,width=4cm,height=6cm,angle=90}
\end{minipage}\hfill\hfill
\begin{minipage}[b]{.495\linewidth}
\centering\epsfig{file=cosbbbb_procs.ps,width=4cm,height=6cm,angle=90}
\end{minipage}
\caption{Differential distributions in invariant mass 
  of multi-jet systems containing two or four $b$ (anti)quarks. 
  Also shown in the bottom right plot is the (cosine of) the angle of
  the 4$b$ system with respect to the beam axis.  The {\sc Cm} energy
  is 500 GeV and the Higgs mass 110 GeV.  Our acceptance cuts in
  energy and separation of the four $b$ quarks have been implemented.}
\label{fig_dist}
\end{figure}

We now investigate several differential spectra, to find kinematic cuts
which will suppress the backgrounds.  The distributions in $E(b)$ and
$\cos(b,b)$ cannot be further exploited after the acceptance cuts are
made.  Instead we consider the invariant masses of $b$ (anti)quark
systems: for 2$b$ systems where the $b$ jets come from the same
production vertex (`right' pairing) or otherwise (`wrong' pairing);
for 3$b$ systems, and for the 4$b$ system.  The spectra for the 2$b$
and 4$b$ systems are shown in Fig.~\ref{fig_dist} (the 3$b$ spectra
are less useful and are not shown here). It is clear that the narrow
Higgs peak\footnote{Recall that for $M_H=110$ GeV one has
  $\Gamma_H\approx3$ MeV.  The Higgs resonances in Fig.~\ref{fig_dist}
  are smeared by incorporating a $5$ GeV bin width.}  in the $M_R(bb)$
distribution, can be exploited to reduce the {\sc Qcd} background.
Although the {\sc Ew} process also displays a resonance at $M_H$, only
one 2$b$ invariant mass will peak here compared to two combinations in
the signal. The {\sc Ew} background can therefore also be cut away.
Finally, one may require that none of the 2$b$ invariant masses
reproduce a $Z$ boson, provided that the invariant mass resolution of
di-jet systems is at least as good as the difference $(M_H-M_Z)/2$, in
order to resolve the $Z$ and $H$ peaks.  The $M(bbbb)$ spectra is also
useful in distinguishing signal from background since clearly
$M(bbbb)$ must always be greater than $2M_H$ for the signal process,
while it can be lower for both background processes (especially for
the {\sc Qcd} background).

In addition to the different resonance structures of signal and
background one can exploit the dominantly $t$-channel nature of the
backgrounds as compared to the $s$-channel signal.  In
Fig.~\ref{fig_dist} we also show the cosine of the polar angle (i.e.
with respect to the beam axes) of the four $b$ quark system (or,
indeed, the real $Z$).  Notice that the backgrounds are much more
forward peaked than the signal.  The {\sc Qcd} events are
predominantly $e^+e^-\to ZZ$ production followed by the decay of one
of the gauge bosons into four $b$ quarks.  The $ZZ$ pair is produced
via $t,u$-channel graphs so the four $b$ quarks are preferentially
directed forwards and backwards into the detector. In contrast, the
signal is entirely $s$-channel resulting in more centrally produced
$b$ jets.  The EW background has a more complicated structure but is
still sizably dominated by forward production.  A similar effect is
seen for $\cos(bbb)$ and $\cos(bb)$, allowing one to separate signal and
backgrounds events efficiently.

The transverse momentum distributions of the 4$b$, 3$b$ and 2$b$
systems were also examined. However the distributions for the signal
and the {\sc Ew} background proved too similar to allow their use as
efficient kinematic variables for cuts.

The main features of distributions studied above are rather stable to
variation of the {\sc Cm} energy or Higgs mass (within the ranges
under discussion). We can therefore optimize the $S/B$ ratio by
imposing the cuts:
$$
|M(bb)-M_H|<5~{\mathrm{GeV}}~({\mathrm{on~exactly~two~combinations~of}}~
2{\it b}~{\mathrm{systems}}),
$$
$$
|M(bb)-M_Z|>5~{\mathrm{GeV}}~({\mathrm{for~all~combinations~of}}~
2{\it b}~{\mathrm{systems}})
$$
\begin{equation}\label{cuts}
M(bbbb)>2M_H,
\qquad\qquad |\cos(2b,3b,4b)|<0.75.
\end{equation}
In enforcing these constraints, we assume no $b$ jet charge
determination.

\begin{figure}[ht]
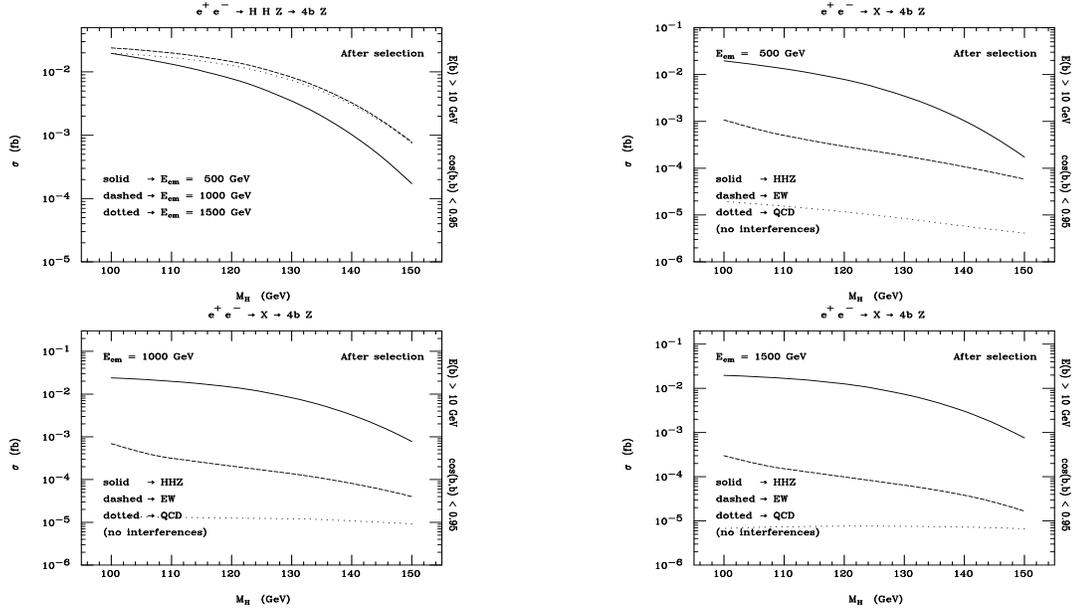

\begin{minipage}[b]{.495\linewidth}
\centering\epsfig{file=signal_cut.ps,angle=90,height=4cm,width=6cm}
\end{minipage}\hfill\hfill
\begin{minipage}[b]{.495\linewidth}
\centering\epsfig{file=background500_cut.ps,angle=90,height=4cm,width=6cm}
\end{minipage}\hfill\hfill
\begin{minipage}[b]{.495\linewidth}
\centering\epsfig{file=background1000_cut.ps,angle=90,height=4cm,width=6cm}
\end{minipage}\hfill\hfill
\begin{minipage}[b]{.495\linewidth}
\centering\epsfig{file=background1500_cut.ps,angle=90,height=4cm,width=6cm}
\end{minipage}
\caption{Top-left: cross sections in femtobarns for the signal at three 
  different collider energies: 500, 1000 and 1500 GeV.
  Top-right(Bottom-left)[Bottom-right]: cross sections in femtobarns
  for the signal versus the {\sc Ew} and {\sc Qcd} backgrounds at
  500(1000)[1500] GeV.  Acceptance and selection cuts have been
  implemented.}
\label{fig_cross_cut}
\end{figure}

The signal and background cross-sections after the implementation of
the selection cuts can be seen in Fig.~\ref{fig_cross_cut}.  Both
background rates are greatly reduced while a large portion of the
original signal is maintained.  This results in a $S/B$ ratio which is
enormously large for not too heavy Higgs masses. For example, at
Centre-of-mass energies $\Ecm=500(1000)[1500]$ GeV and for $M_H=110$
GeV, one finds $S/B=25(60)[104]$.  This remarkable suppression of the
backgrounds comes largely from the invariant mass cuts on $M_{bb}$.
In fact, they are crucial not only in selecting the $M_H$ resonance of
the signal, but also in minimizing the signal rejection around $M_Z$
when mispairings occur: notice the shoulder at 90 GeV of the $M_W(bb)$
signal spectrum.

A number of caveats to our analysis apply. Firstly, the value we have
adopted for the resolution is rather high, considering the large
uncertainties normally associated with the experimental determination
of jet angles and energies, though not unrealistic in view of the most
recent studies \cite{resolution}. The ability of the actual detectors
in guaranteeing the performances foreseen at present is thus crucial
for the feasibility of dedicated studies of double Higgs-strahlung
events at the LC.  Furthermore, one must consider the efficiency of
tagging the $b$ quarks necessarily present in the final state,
particularly in the case in which the $Z$ boson decays hadronically.
Given the high production rate of six jet events from {\sc Qcd}
\cite{ee6j} and multiple gauge boson resonances \cite{gauge} in light
quark and gluon jets, it is desirable to resort to heavy flavour
identification in hadronic samples. However, the poor statistics of
the $HHZ$ signal requires a judicious approach in order not to deplete
it below detection level. According to recent studies \cite{btag}, the
two instances can be combined successfully, as efficiencies for
tagging $b\bar b$ pairs produced in Higgs decays were computed to be
as large as $\epsilon_{b\bar b}\approx90\%$, with mis-identification
probabilities of light(charmed) quarks as low as $\epsilon_{q\bar
  q(c\bar c)}\approx0.3(4)\%$ (and negligible for gluons).  If such a
projection for the {\sc Lc} detectors proves to be true, then even the
requirement of tagging exactly four $b$ quarks in
double Higgs-strahlung events might be statistically feasible, thus
suppressing the reducible backgrounds to really marginal levels
\cite{Lutz}.  One should also bear in mind that experimental
considerations, such as the performances of detectors, the
fragmentation/hadronization dynamics and a realistic treatment of the
$Z$ boson decays, are also important when determining what cuts should
be made. Such considerations are beyond the scope of this paper, and
are under study elsewhere \cite{Lutz}.

Finally, the number of signal and backgrounds events seen per inverse
attobarn of luminosity at $\Ecm=500$, $1000$, and $1500$ GeV, with
$M_H=110$ GeV, can be seen in Tab.~\ref{eventtable}. One could relax
one or more of the constraints we have adopted to try to improve the
signal rates without letting the backgrounds become unmanageably
large.  However, such a relaxation only marginally effects the signal
while significantly reducing the $S/B$ ratio, and should only be done if
high luminosities cannot be obtained. Kinematic fits can also help in
improving the $S/B$ ratio \cite{Lutz}.
\begin{table}[ht]
\begin{center}
\begin{tabular}{|c||c|c|c|}
\hline
& \multicolumn{3}{c|}{Number of Events per $ab^{-1}$ after selection 
cuts}  \\ \cline{2-4}
& $E_{\rm cm}=500$ GeV & $E_{\rm cm}=1000$ GeV & $E_{\rm cm}=1500$ GeV \\ \hline\hline
signal & $26$ & $40$ & $34$ \\ \hline
Electroweak& $1.0$ & $0.6$ & $0.3$ \\ \hline
{\sc Qcd} & $0.032$ & $0.026$ & $0.016$ \\ \hline
\end{tabular}
\caption{The number of events for signal and backgrounds per 
  inverse attobarn of luminosity after selection cuts for
  centre-of-mass energies of $500$, $1000$ and $1500$ GeV, a Higgs
  mass of $110$ GeV, and with polarized electron and positron beams.}
\label{eventtable}
\end{center}
\end{table}

\vspace{-1cm}

\section{Summary}
\label{sec_conclusions}

In conclusion, the overwhelming irreducible background from {\sc Ew}
and {\sc Qcd} processes of the type $e^+e^-\to b\bar b b\bar b Z$ to
double Higgs production in association with $Z$ bosons and decay in
the channel $H\to b\bar b$, i.e., $e^+e^-\to HHZ\to b\bar b b\bar b
Z$, should easily be suppressed down to manageable levels by simple
kinematics cuts: e.g. in invariant masses and polar angles.

The number of signal events is generally rather low, but will be
observable at the {\sc Lc} provided that it provides very high
luminosity, excellent $b$ tagging performances, and high di-jet
resolution.  As advocated in Ref.~\cite{Lutz}, one also requires a
good forward acceptance for jets since single jet directions in the
double Higgs-strahlung process can stretch up to about $20^{\rm
{\small o}}$ in polar angle.




\begin{thebibliography}{99}
  
\vspace{-0.1cm}

\bibitem{ee500} Proceedings of the Workshop {\it $e^+e^-$ Collision
    at 500 GeV. The Physics Potential}, Munich, Annecy, Hamburg, 3--4
  February 1991, ed.~P.M.~Zerwas, DESY 92--123A/B,
  August 1992, DESY 93--123C, December 1993; \\
  E. Accomando et al., {\it Phys. Rep. Vol. {\bf 299}}, (1998) 1. 

\bibitem{GlMa} E.W.N.~Glover and A.D.~Martin, {\it Phys. Lett.} {\bf
    B226} (1989) 393; \\
  W.~Bernreuther, M.J.~Duncan, E.W.N.~Glover, R.~Kleiss, J.J.~van der
  Bij, J.J.~Gomez-Cadenas and C.A.~Heusch, CERN-TH-5484-89,
  contribution to proceedings of the Workshop ``Z Physics at LEP'',
  LEP Phys. Wrkshp. v.2 (1989) 1-57.

\vspace{-0.1cm}

\bibitem{PMZ} A.~Djouadi, W.~Kilian, M.~M\"uhlleitner and P.M.~Zerwas,
  {\it Eur. Phys. J.} {\bf C10} (1999) 27; contribution to the XXIX
  {\em International Conference on High Energy Physics}, Vancouver
  1998, Heidelberg Report HD-THEP 98-29.
  
\vspace{-0.1cm}

\bibitem{MMM} R.~Lafaye, D.~J.~Miller, S.~Moretti and
  M.~M\"uhlleitner, contribution to proceedings of workshop on
  ``Physics at TeV Colliders'', Les Houches, France 8-18 June 1999.
  
\vspace{-0.1cm}

\bibitem{US} D.~J.~Miller and S.~Moretti, June 1999, {\tt hep-ph}{\tt
    /9906395}.
  
\vspace{-0.1cm}

\bibitem{ISR} T.~Barklow, P.~Chen and W.~Kozanecki, in
  Ref.~\cite{ee500}, part A.
  
\vspace{-0.1cm}

\bibitem{WJSZK} Z.~Kunszt, S.~ Moretti and W.J.~ Stirling, {\it
    Z.~Phys.} {\bf C74} (1997) 479.

\vspace{-0.1cm}

\bibitem{resolution} F. Richard, private communication.
  
\vspace{-0.1cm}

\bibitem{ee6j} S. Moretti, {\it Phys. Lett.} {\bf B420} (1998) 367;
  {\it Nucl. Phys.} {\bf B544} (1999) 289.

\vspace{-0.1cm}

\bibitem{gauge} A. Aeppli {\it et. al.}, in Ref.~\cite{ee500}, part A;\\
  V. Barger and T. Han, \pl B212 1988 117;\\
  V. Barger, T. Han and R.J.N. Phillips, \pr D39 1889 146;\\
  A. Tofighi-Niaki and J.F. Gunion, {\it Phys. Rev.} {\bf D39} (1989)
  720.

\vspace{-0.1cm}

\bibitem{btag} G. Borissov, talk delivered at the ECFA/DESY Workshop
  on ``Physics and Detectors for
  a Linear Collider'', Oxford, UK, March 20--23, 1999;\\
  M. Battaglia, {\it ibidem}.

\vspace{-0.1cm}

\bibitem{Lutz} P. Lutz, talk delivered at the ECFA/DESY Workshop on
  ``Physics and Detectors for a Linear Collider'', Oxford, UK, March
  20--23, 1999.

\end{thebibliography}
\end{document}